\documentstyle[aps,prl,multicol,epsfig]{revtex}
\renewcommand{\v}[1]{{\bf #1}}
\newcommand{\s}{{\sigma}}

\newcommand{\Psib}{{\bar{\Psi}}}
\newcommand{\bb}{{\bar{b}}}

\newcommand{\sign}{{\rm sign}}

\newcommand{\eq}{\begin{equation}}
\newcommand{\ee}{\end{equation}}

\newcommand{\p}{{\partial}}
\newcommand{\gr}{{\nabla}}

\def\eqa{\begin{eqnarray}}
\def\eea{\end{eqnarray}}

\newcommand{\nn}{\nonumber\\}
\newcommand{\Eq}[1]{Eq.~(\ref{#1})}

\newcommand{\up}{\uparrow}
\newcommand{\down}{\downarrow}

\newcommand{\al}{\alpha}

\newcommand{\eps}{\epsilon}

\begin{document}
\draft
\widetext

\title{Pairing via Index theorem}

\author{Dung-Hai Lee}
\maketitle

\widetext
\begin{abstract}
\rightskip 54.8pt

This work is motivated by a specific point of view: at short distances and high energies the undoped and
underdoped cuprates resemble the $\pi$-flux phase of the t-J model. The purpose of this paper is to present a
mechanism by which pairing grows out of the doped $\pi$-flux phase. According to this mechanism pairing symmetry
is determined by a parameter controlling the quantum tunneling of gauge flux quanta. For zero tunneling the
symmetry is $d_{x^2-y^2}+id_{xy}$, while for large tunneling it is $d_{x^2-y^2}$. A zero-temperature critical
point separates these two limits.
\end{abstract}


\begin{multicols}{2}

\narrowtext

It is now well appreciated that the unconventional properties of the cuprate superconductors are best manifested
in the underdoped regime. For example there is a crossover temperature $T^*>T_c$ below which a pseudo spin gap
opens up.  In addition, recent angle-resolved photoemission spectroscopy shows a common feature between
the underdoped and undoped cuprates  - a high-binding-energy peak whose dispersion is similar to that of the
spinon in the $\pi$ flux phase\cite{affleck} of the half-filled t-J Model.\cite{shen} Due to this result, there is
a spreading feeling that at short distances and high energies the undoped and underdoped cuprates resemble the
$\pi$ flux phase of the antiferromagnet.\cite{hsu,bob,wl,kl}

For undoped cuprates this feeling is supported by a recent work of Kim and Lee.\cite{kl}
Starting from the free Dirac
spinons of the $\pi$ flux phase, Kim and Lee show that gauge fluctuation binds spinon and antispinon and causes their condensation
into the ordered Neel state at low temperatures. This work demonstrates that even though the $\pi$ flux phase
physics (i.e. free spinon excitations) is distant from the reality, gauge fluctuations can restore the right low
temperature behaviors.

In the following we extend this point of view to the underdoped regime. The question we would like to ask is ``can
gauge fluctuation produce pairing out of the doped $\pi$-flux phase''. Interestingly we find a particular
mechanism which yields either $d_{x^2-y^2}+id_{xy}$ or $d_{x^2-y^2}$ pairing symmetries depending on a parameter
controlling the quantum tunneling of gauge flux quanta from $+\phi_0$ to $-\phi_0$.
\\

\noindent{\bf{The model:}} \rm
\\

The model we shall consider is the U(1) gauge theory of the doped $\pi$-flux phase. The low energy degrees of freedom
are: spinons (Dirac-like fermions) , holons (non-relavistic bosons), and an U(1) gauge field. The model is defined
by: \eq Z=\int D[\bar{\Psi}_{\pm},\Psi_{\pm}]D[\bb,b]D[a_{\mu}]e^{-\int dtd^2x{\cal L}},\ee where \eqa {\cal L}&&=
\bb~[(\p_t-\mu_b-ia_0)+\frac{1}{2m_b}(\v p-\v a)^2]~b+~\frac{g}{2}|\bb b|^2 \nn &&+ \sum_{n=\pm}
\Psib_{n,\al}[(\p_t-\mu_f-ia_0)+ v \v{{\s}}\cdot(\v p-\v a)]\Psi_{n,\al}.\label{dirac1}\eea In the above $\pm$
refers to the two ``Dirac points'' $\v Q_{\pm}\equiv (\frac{\pi}{2a},\pm\frac{\pi}{2a})$; $\al=\up,\down$ labels
the spin state; $\Psi_{\pm\al}=\pmatrix{\psi_{1\pm\al}\cr \psi_{2\pm\al}\cr}$ are spinon fields; $b$ is the holon
field; $\v p\equiv \gr/i$; and $\s_x,\s_y,\s_z$ are Pauli matrices.

The fluctuation of $a_{\mu}$ tends to bind particles of opposite gauge charges. There are two obvious candidates
for the bound states: (a) a spinon-antispinon pair, and (b) a holon-antispinon pair. In the undoped
antiferromagnetic phase, the first scenario is realized.  Indeed, the Neel-ordered state can be thought of as a
condensate of the spinon-antispinon (i.e. magnon) pairs.
We view the unusual properties of the underdoped regime as due to the competition and compromise between binding
scenarios (a) and (b). A particular example of the compromise is the formation of stripes.\cite{steve} In this
case holon-antispinon and spinon-antispinon exist in different spatial regions. Another example is the formation
of pairs of physical holes. In this case two holons and two antispinons (hence (a) and (b)) occur in the same
composite.  This pairing mechanism is the subject of the rest of the paper.
\\

\noindent{\bf{{The index theorem}}
\\
\rm

Let $\v a(\v x)$ be a static gauge field. The eigenvalue problem \eq \v {{\s}}\cdot(\v p-\v
a)\Psi=E\Psi\label{indx} \ee has $N_{\phi}-1$ normalizable zero-energy solutions, where $N_{\phi}=\int
d^2x(\gr\times\v a(\v x))/\phi_0$.\cite{ac}
\\

\noindent{\bf{{Pairing by index theorem}}
\\
\rm

Let us imagine starting from the half-filled $\pi$-flux phase, and allow $\gr\times\v a$ to nucleate $N+1$ flux
quanta. By index theorem, such flux quanta produce $4N$ zero-energy spinon levels. ($4N$, because of two Dirac
points and two spin directions.) At half-filling $2N$ of these levels are occupied. As the result the flux quanta
lift $2N$ spinons to the Fermi energy so that they can be removed by doping. We note that in the above argument
there is a free choice, namely, the polarity of the flux quanta.

Now let us come to finite doping ($x>0$). In order to produce the right number of zero-energy spinons
we have to allow a density, $x/2$, of flux quanta in  $\gr\times\v a$. In order for these spinons to be true
zero-energy these flux quanta must have the same polarity.

In the above discussion the flux quanta are implicitly assumed to be static. In the following we shall relieve
that assumption. We shall treat $\gr\times\v a$ as a fluctuating flux density satisfying \eq
<\gr\times\v a(\v r,t)>=B=\frac{x}{2}\phi_0.\ee (Here $<...>$ stands for space and time average.)
At the mean-field level, the spinons see an uniform magnetic field, and have the following spectrum \eq
E_m=\sign(m)v\sqrt{2B}\sqrt{|m|}.\label{landau}\ee At doping concentration $x$ the $E=0$ Landau band is empty and
all bands below that are occupied for both spices of spinons. Consequently, single spinon excitation is gapped.

Now let us go beyond mean-field. We write \eq a_{\mu}=a_{ext,\mu}+\delta a_{\mu}, \ee where $\gr\times\v
a_{ext}(\v r,t)=\frac{x}{2}\phi_0$ and $a_{ext,0}=0$. By integrating out the spinon excitations above the
mean-field vacuum, we obtain the following effective action for $\delta a_{\mu}$: \eq S_{eff}=\int dtd^2x
\{\frac{1}{2\kappa_1}|\v e|^2 + \frac{1}{2\kappa_2}b^2 -\frac{i}{2\pi}\eps_{\mu\nu\lambda}\delta
a_{\mu}\p_{\nu}\delta a_{\lambda}\}. \label{gs} \ee Here $\v e\equiv (\p_0\delta a_1-\p_1\delta a_0,\p_0\delta
a_2-\p_2\delta a_0)$ and $b\equiv \p_1\delta a_2-\p_2\delta a_1$ are the electro-magnetic field associated with
$\delta a_{\mu}$. \Eq{gs} implies that the long wavelength spinon density fluctuation ($\delta\rho_{spinon}$) is
locked to the fluctuation of $\gr\times\delta\v a$ so that \eq
2\delta\rho_{spinon}=\frac{1}{\phi_0}\gr\times\delta\v a. \label{lock}\ee Physically \Eq{lock} implies that two
antispinons are bound to a quantum of $\gr\times\v a$.

Next we look at the holons. By putting \Eq{gs} together with the remaining boson terms in \Eq{dirac1} we obtain:
\eqa S_b&&=\int dt d^2x\{\bb[(\p_t-i\delta a_0-\mu_b) -\frac{1}{2m_b}(\gr-i\v a_{ext}-i\delta\v a)^2]b\nn
&&+\frac{1}{2\kappa_1}|\v e|^2 + \frac{1}{2\kappa_2}b^2 -\frac{i}{2\pi}\eps_{\mu\nu\lambda}\delta
a_{\mu}\p_{\nu}\delta a_{\lambda}\}. \label{dirac2}\eea \Eq{dirac2} describes semions in an external magnetic
field $\gr\times \v a_{ext}$ corresponding to filling factor $\nu=2$. For semions this is the right filling factor
for the formation of quantum Hall liquid. Thus at low energies \eq
2\delta\rho_{holon}=\frac{1}{\phi_0}\gr\times\delta\v a.\ee Physically this means that two holons bind with each
quantum in $\gr\times\v a$.

Thus each quantum of $\gr\times\v a$ binds two antispinons and two holons together.
The composite is a pair of
physical holes. The pairing symmetry is $d_{x^2-y^2}+id_{xy}$, exactly the same as that obtained in
Ref.\cite{anyon,rokhsar}. In addition this pairing mechanism is one
example of the topological pairing of Wiegmann.\cite{wiegmann}
\\

\noindent{\bf{Instanton effects}}
\\
\rm

It is likely that we are now loosing readers. This is because there is a consensus that the T-breaking pairing
mentioned above is not realized in the cuprates. In the following we demonstrate a mechanism by which the
$d_{x^2-y^2}+id_{xy}$ pairing state is turned into a pure $d_{x^2-y^2}$ one.

The mechanism is the quantum tunneling of the $\gr\times\v a$  quanta from  $+\phi_0$ to $-\phi_0$ (or vice
versa). In continuum space, such event (often referred to as the instanton) requires singularity in the gauge
field configuration. There is no such need on a lattice. Moreover since the total gauge charge (i.e. the spinon +
holon number) per site is always 1, such event does not cost diverging action. For example a finite-action
instanton event can occur by first shrinking the size of a flux quantum until it is contained in a single
plaquette, then reverse it polarity, i.e.

\begin{enumerate}
\item shrink a flux quantum to one plaquette,
\item reverse the polarity of the flux quantum,
\item expand the reversed flux quantum to the original size.
\end{enumerate}

An effect of such tunneling is to produce quantum mixing between the $d_{x^2-y^2}+id_{xy}$ and
$d_{x^2-y^2}-id_{xy}$ pairing states, namely, \eq (d_{x^2-y^2}+id_{xy})+(d_{x^2-y^2}-id_{xy})\sim d_{x^2-y^2}.\ee
\\

\noindent{\bf{T-breaking versus non T-breaking}}
\\
\rm

Whether the overall pairing break time reversal symmetry depends on the competition between the following
energetics. First, let us ignore the quantum tunneling of gauge flux. Under such condition the minimum energy is
obtained when the polarity of the flux quanta order ferromagnetically. The reversal of a single
flux quantum costs an energy
equals to that of breaking two $d_{x^2-y^2}+id_{xy}$ pairs. Such energy cost is reminiscent to the spin-flip energy
of an Ising ferromagnet.

In this analogy instanton acts as a transverse field on the Ising spins.
Thus the model describing the dynamics of
gauge flux is a transverse-field-Ising-like model:
\eq H=-K\sum_{<ij>}\s^z_i\s^z_j-h\sum_i\s^x_i. \label{phe1} \ee
In \Eq{phe1} $K$ prefers T-breaking while $h$ tends to restore T.
The parameter that determines time-reversal
symmetry breaking is $R=\frac{K}{h}.$ For $R>R_c$
T-breaking will occur, and for $R<R_c$
time reversal remains unbroken. The quantum critical point separating these two phases
is where the chirality fluctuation diverges.

If the above model is applicable to the high $T_c$ superconductors, then the
$K/h$ for the cuprates must satisfy
$R<R_c$.

In the above we have argued that the severe fluctuation (dominated by the instanton events) of $\gr\times\v a$
favors the formation of $d_{x^2-y^2}$ Cooper pairs. It is important to stress that each Cooper pair is made up of
two physical holes (i.e. two holons and two antispinons) carrying zero $a_{\mu}$-gauge charge. Because of that the
Cooper pair motion is not affected by the strong gauge fluctuations. This decoupling from the gauge field is
essential for their ability to condense into a superconducting state. Above the superconducting transition the
Cooper pairs persist to exist.\cite{ek} It is physically clear that in the present scenario the formation of
quasiparticle gap is not driven by the condensation of Cooper pairs as in the BCS theory.  In addition, because
the charge-flux bound state has a algebraic profile  in the index theorem, the present pairing mechanism is unlike
the ``molecular'' limit where holes are tight into small real-space pairs.
\\

\noindent{\bf{Disclaimer}}
\\
\rm

Obviously this work does not represent an unbiased solution of a pecific microscopic model. Moreover its starting
point (i.e. the $\pi$ flux phase) is based on a point of view whose validity
is not commonly accepted.\cite{but}
\\

\noindent{\bf{Acknowledgement}}
\\
\rm
I thank A. Balatsky and M. Greiter and D.H. Kim for discussions, and J.H. Han for
pointing out Ref.\cite{wiegmann}. I also thank R.B Laughlin for pressing the
importance of the question ``why the $\pi$ flux phase?''. Some
discussions on this issue can be found in Ref.\cite{laughlin3}.


}}
\end{multicols}


\begin{thebibliography}{99}
\bibitem{but} So far we focused on the doped $\pi$ flux phase. Similar arguments can also be applied to the staggered flux
phase of Wen and Lee\cite{wl}. For the staggered flux phase there are two spices of boson instead of one. These
two kinds of boson carry opposite gauge charge, and have equal density ($x/2$) in the mean-field vacuum. In the
spirit discussed above each quantum of gauge flux converts a ``$b_2$ boson'' into a $b_1$ type,\cite{wl} and binds
two $b_1$ bosons with two antispinons.
\end{thebibliography}
\end{document}